\documentstyle[12pt]{article}
\begin{document}
\baselineskip 20pt
\begin{titlepage}
\begin{flushright}{FAU-TP3-00/1}
\end{flushright}
\vskip 3.0cm
\begin{center}
{\Large {\bf Decompactification of space or time in large $N$ QCD$_2$}}
\vskip 0.5cm
Verena Sch\"on and Michael Thies
\vskip 0.2cm
{\it Institute for Theoretical Physics III,
University of Erlangen-N\"urnberg, \\
 Staudtstr. 7, 91058 Erlangen, Germany}
\end{center}
\vskip 3.0cm
\begin{abstract}
QCD$_2$ with fundamental quarks on a cylinder is solved to leading order
in the $1/N$ expansion, including the zero mode gluons. As a result of the
non-perturbative dynamics of these gauge degrees of freedom, the compact
space-time direction
gets effectively decompactified. In a thermodynamic interpretation, this
implies that there is no pressure of order $N$ and that the chiral condensate
of order $N$ is temperature independent. These findings are consistent
with confinement of quarks, rule out both
chiral and deconfining phase transitions in the finite temperature
't~Hooft model, and help to resolve some
controversial issues in the literature.
\end{abstract}

\end{titlepage}

Whereas 2-dimensional field-theoretic models like the Schwinger model
\cite{Schwinger}
or the Gross-Neveu model
\cite{Gross-Neveu}
are by now well understood also at finite
temperature
\cite{Sachs-Wipf,HYDMR},
the situation is less clear in the case of the 't~Hooft
model \cite{tHooft}, large $N$ QCD$_2$ with fundamental quarks.
In view of the similarity
between these models, notably their chiral aspects, and the vast amount
of literature on QCD$_2$, this is rather surprising. Briefly, the
present situation is as follows: McLerran and Sen \cite{McLerran-Sen} argue
that there is no deconfining phase transition, except possibly at
infinite temperature.
Ming Li \cite{Ming Li}, using standard finite temperature field theory
methods, concludes that chiral symmetry may get restored in
the limit $T\to \infty$.
In both of these studies, severe infrared problems were encountered,
either in the form of divergent diagrams or ambiguous quark
self energies. In a different vein, several studies have addressed
QCD$_2$ on a spatial circle at zero temperature. By interchanging
Euclidean time with space and invoking covariance, this can also be
reinterpreted as finite
temperature calculations for a spatially extended system, even though the
techniques
used are quite different. Lenz {\em et al.} \cite{LTLY}
observe a chiral phase transition
in the massless 't~Hooft model
at some critical length, strongly reminiscent of the Gross-Neveu model.
Dhar {\em et al.} \cite{DLMW} treat the zero mode gluons in a more
ambitious way
than ref. \cite{LTLY}, using technology from matrix models and string
theory, but are not able to fully solve the resulting complicated equations.
They propose that the gauge variables get decompactified by the
fermions in complete analogy with the Schwinger model
\cite{Manton},
a claim which
has recently been disputed by Engelhardt
\cite{Engelhardt}.

In view of this unsatisfactory state of the art, we have reconsidered the
't~Hooft model on a cylinder and, in particular, tried to clarify the
role of the zero mode gluons. We shall present here a novel treatment
of the leading order in the large $N$ expansion, which seems to resolve
the above sketched discrepancies. This opens the way to a controlled study
of the $1/N$ corrections expected to reveal the true, ``hadronic"
physics of the model.

We work canonically on a spatial circle of length $L$ in the gauge
$\partial_1 A_1 =0$, $(A_1)_{ij}=\delta_{ij}\frac{\varphi_i}{gL}$ diagonal in
color. The Hamiltonian reads (cf. \cite{LTLY,LNT,LST})
\begin{equation}
H=H_{\rm g} + H_{\rm f} + H_{\rm C} \ ,
\label{p1}
\end{equation}
with the gauge field kinetic energy
\begin{equation}
H_{\rm g}=-\frac{g^2 L}{4} \sum_i \frac{\partial^2}{\partial \varphi_i^2}
\ ,
\label{p2}
\end{equation}
the quark kinetic energy
\begin{equation}
H_{\rm f}  = \sum_{n,i} \frac{2\pi}{L}\left( n+\frac{\varphi_i}{2\pi}\right)
\left( a_i^{\dagger}(n)a_i(n)
- b_i^{\dagger}(n)b_i(n) \right)
+m \sum_{n,i} \left(a_i^{\dagger}(n)b_i(n)+b_i^{\dagger}(n)
a_i(n)\right) \ ,
\label{p3}
\end{equation}
and the Coulomb interaction
\begin{equation}
H_{\rm C} = \frac{g^2 L}{16 \pi^2} \sum_{n,i,j} \frac{j_{ij}(n)j_{ji}(-n)}
{\left(  n-\frac{\varphi_j-\varphi_i}{2\pi}\right)^2} \ .
\label{p4}
\end{equation}
Here, the $a_i(n), b_i(n)$ denote second quantized right- and left-handed
quarks,
respectively, and the currents $j_{ij}(n)$ can be taken in the U($N$) form
at large $N$,
\begin{equation}
j_{ij}(n)=\sum_{n'} \left( a_j^{\dagger}(n') a_i(n'+n) + b_j^{\dagger}(n')
b_i(n'+n) \right) \ .
\label{p5}
\end{equation}
As explained elsewhere \cite{LNT,LST}, due to the curved
configuration space of the $\varphi_i$ and the SU($N$)
Haar measure originally appearing in $H_{\rm g}$, this Hamiltonian has to be
supplemented by the following boundary condition for the
wavefunctionals,
\begin{equation}
\Psi(\varphi_1,...,\varphi_N;{\rm fermions}) = 0 \quad {\rm if} \quad
\varphi_i=\varphi_j \ {\rm mod}\  2\pi \ .
\label{p5a}
\end{equation}

In ref. \cite{LTLY}, quantization was performed after complete
classical gauge fixing.
In this way, the fact that the $\varphi_i$ are curvilinear coordinates
is missed.
This led to the assumption that all the
$\varphi_i$ are frozen at the value $\pi$ in the large $N$ limit. In the
resulting
purely fermionic theory, the only remnant of the gluons are
antiperiodic boundary conditions for the quarks in the compact space
direction. In the meantime,
this whole approach has been put on a more rigorous basis by first
quantizing in the Weyl gauge
and then resolving the Gauss law quantum mechanically \cite{LNT}. This
made it clear that the
$\varphi_i$ are parameters on the group manifold
with corresponding Jacobian, the SU($N$) reduced Haar measure.
When solving the theory, it is then possible to restrict
oneself to
the smallest region in field space bounded by zeros of the Jacobian, cf.
ref. \cite{LST} where
the consequences for SU(2), SU(3) where explored in the case of
adjoint fermions.
How can this be
generalized to the large $N$ limit?
A definite choice of ``fundamental
domain" obviously means that the $\varphi_i$ always remain ordered, say
$0 \leq \varphi_1 \leq \varphi_2 \leq ... \leq \varphi_N \leq 2\pi$.
If we think of the gluons as particles on a circle, they cannot
cross each other and become closely packed in the limit $N\to \infty$.
Their fluctuations are suppressed by $1/N$, simply due to lack
of space. The only
degree of freedom left, the collective rotation of this ``pearl
necklace", is a U(1) factor which anyway is not present
in the SU($N$) theory. This
suggests that the correct choice for the gluon background field
as seen by the fermions
is not $\varphi_i=\pi$, but rather
the continuum limit of the lattice points
\begin{equation}
\varphi_i = 2\pi \frac{i}{N} \ , \quad i=1...N \ .
\label{p6}
\end{equation}
Instead of antiperiodic boundary conditions, the fermions then acquire
color dependent boundary conditions which interpolate smoothly
between the phases 0 and $2\pi$,
\begin{equation}
\psi_k(L)={\rm e}^{{\rm i}2\pi k/N}\psi_k(0) \ , \quad (k=1,...,N) \ .
\label{p6a}
\end{equation}
Note that in such a completely gauge fixed formulation, there is
nothing wrong with having a color dependence of this type.

In the thermodynamic limit $L\to \infty$, both of these choices
of the gluon field configuration, $\varphi_i = \pi$ or eq. (\ref{p6}),
will become indistinguishable and yield
the well-known results.
We now show that at finite $L$, the effects of the gluons on the
quarks is radically different in these two cases, and that it is the
spread out
distribution (\ref{p6}) which is in fact the correct one.

The fermions can be treated in a relativistic Hartree-Fock approximation
along the lines explained in ref. \cite{LTLY}. In this approach, all the
information about the vacuum is encoded in the Bogoliubov angle
$\theta_i(n)$, related to self-consistent Hartree-Fock spinors via
\begin{equation}
u_i(n) = \left(\begin{array}{c} \cos \theta_i(n)/2 \\ \sin \theta_i(n)/2
\end{array} \right) \ , \quad
v_i(n) = \left(\begin{array}{r} -\sin \theta_i(n)/2 \\ \cos \theta_i(n)/2
\end{array} \right) \ .
\label{p7}
\end{equation}
The Bogoliubov angles in turn are determined by the gap equation,
\begin{equation}
\frac{2\pi}{L}(n+\alpha_i) \sin \theta_i(n) - m \cos \theta_i(n)
+ \frac{g^2L}{16 \pi^2} \sum_{n',j}\frac{\sin \left( \theta_i(n)-
\theta_j(n-n')\right)}{(n'-\alpha_j+\alpha_i)^2} = 0 \ .
\label{p8}
\end{equation}
Here, we have switched to the slightly more convenient variable
$\alpha_i = \frac{\varphi_i}{2\pi}
\in [0,1]$ for the gluons.
If $\alpha_i=1/2$ as chosen in ref. \cite{LTLY}, $\theta_i(n)$ becomes
$i$-independent and we recover the old gap equation considered in that work.
Now, we assume $\alpha_i=i/N$ and perform the large $N$ limit before
solving the gap equation. Since $\alpha_i$ becomes a continuous variable,
we replace
$\theta_i(n)\to \theta_{\alpha}(n)$ and
$\sum_j \to N \int_0^1 {\rm d}\alpha'$, with the result
\begin{equation}
\frac{2\pi}{L}(n+\alpha) \sin \theta_{\alpha}(n) -m\cos \theta_{\alpha}(n)
+ \frac{Ng^2L}{16\pi^2}
\sum_{n'} \int_0^1 {\rm d} \alpha' \frac{\sin \left( \theta_{\alpha}(n)
-\theta_{\alpha'}(n-n')\right)}{(n'-\alpha'+\alpha)^2} = 0 \ .
\label{t3}
\end{equation}
This infinite set of coupled integral equations collapses into a single,
one-dimensional
integral equation, if we set
\begin{equation}
\theta_{\alpha}(n)=\theta(n+\alpha) \ .
\label{t4}
\end{equation}
Since $n$ is integer and $\alpha \in [0,1]$, this step in effect
decompactifies
the original spatial circle. With this ansatz, the notation
$n+ \alpha= \nu$, $n-n'+\alpha'=\nu'$ (where $\nu, \nu'$ are dimensionless,
continuous
variables) and the substitution $\sum_{n'} \int_0^1 {\rm d}\alpha'
\to \int_{-\infty}^
{\infty} {\rm d} \nu'$, we obtain
\begin{equation}
\frac{2\pi}{L}\nu \sin \theta(\nu) -m\cos\theta(\nu)
+ \frac{Ng^2L}{16\pi^2}\int
\!\!\!\!\!\!- {\rm d}\nu'
 \frac{\sin (\theta(\nu)-\theta(\nu'))}{(\nu
-\nu')^2} = 0 \ .
\label{t5}
\end{equation}
Guided by what is known from the 't~Hooft model in the limit
$L\to \infty$, we have defined
the integral as principal value integral. After rescaling
the variables via
\begin{equation}
\frac{2\pi}{L} \nu := p \ , \ \ \frac{2\pi}{L} \nu' :=p' \ ,
\label{t6}
\end{equation}
where $p, p'$ have the dimension of momenta, we recover exactly
the continuum version
of the Hartree Fock equation, namely
\begin{equation}
p \sin \theta \left( \frac{Lp}{2\pi} \right) -m \cos \theta
\left(\frac{Lp}{2\pi}\right)
+ \frac{Ng^2}{4}
\int\!\!\!\!\!\!- \frac{{\rm d} p'}{2\pi} \frac{\sin \left( \theta
\left( \frac{Lp}{2\pi} \right)
-\theta \left( \frac{Lp'}{2\pi} \right) \right)}{(p-p')^2} = 0 \ .
\label{t7}
\end{equation}
Denoting the Bogoliubov angle of the continuum 't~Hooft model by
$\theta_{\rm cont}(p)$,
we conclude that
\begin{equation}
\theta(\nu) = \theta_{\rm cont}\left(\frac{2\pi}{L}\nu \right) \ ,
\label{t8}
\end{equation}
or, in terms of the original, color dependent Bogoliubov angle,
\begin{equation}
\theta_i(n) \approx \theta_{\rm cont}\left(
\frac{2\pi}{L} \left(n+\frac{i}{N}\right) \right) \ , \quad (N\to \infty) \ .
\label{t9}
\end{equation}
This last relation becomes exact in the limit $N \to \infty$ only. In this
limit, the color-
and $L$-dependent Bogoliubov angles for the 't~Hooft model on the circle
of length $L$
are all given by one universal function,
namely the momentum dependent Bogoliubov angle of the 't~Hooft model
on the infinite line. We emphasize that this universality only holds
for the ``pearl necklace" type distribution of gauge variables,
eq. (\ref{p6}).
If the $\varphi_i$
are all set equal to $\pi$, there is no analytic way known how to relate
$\theta(n)$ for different $L$ values, but one has to solve the gap
equation numerically for each $L$ \cite{LTLY}.

The upshot of this simple exercise is the following: In the large $N$ limit,
the gluon variables influence the fermion boundary conditions in such
a way that the circle gets replaced by a line; they decompactify
space-time. The length $L$ of the spatial circle becomes an irrelevant
parameter.
To confirm this last point, let us evaluate
those
ground state expectation values which are of interest for the bulk
thermodynamic properties
of the system, if one interchanges $L$ and
$\beta=1/T$.
This can be done most conveniently with the help of the key relations
\begin{eqnarray}
\langle a_i^{\dagger}(n)a_i(n')\rangle & = & \frac{1}{2}\delta_{n,n'}
 \left(1-\cos
\theta_i(n)\right) \ , \nonumber \\
\langle b_i^{\dagger}(n)b_i(n')\rangle & = & \frac{1}{2} \delta_{n,n'}
\left(1+\cos
\theta_i(n)\right) \ , \nonumber \\
\langle a_i^{\dagger}(n)b_i(n')\rangle & = &
\langle b_i^{\dagger}(n)a_i(n')\rangle \ = \ -\frac{1}{2} \delta_{n,n'}
\sin \theta_i(n) \ .
\label{t9a}
\end{eqnarray}
The vacuum energy density is given by
\begin{eqnarray}
{\cal E}_{\rm vac} &=& -\frac{1}{L}\sum_{n,i} \left(\frac{2\pi}{L}(n+
\alpha_i)\cos \theta_i(n)
+m \sin \theta_i(n)\right) \nonumber \\
& & + \frac{g^2}{32 \pi^2} \sum_{n,n',i,j} \frac{1-
\cos (\theta_i(n)-\theta_j(n-n')}{(n'-\alpha_j+\alpha_i)^2} \ .
\label{t10}
\end{eqnarray}
It corresponds to the negative of the pressure, in the other
picture.
Using the same substitutions as above, ${\cal E}_{\rm vac}$
can be converted
into the standard continuum expression,
\begin{eqnarray}
{\cal E}_{\rm vac} & = &- \frac{N}{L} \sum_n \int_0^1 {\rm d}\alpha
\left(\frac{2\pi}{L}(n+\alpha) \cos \theta(n+\alpha) +m \sin \theta(
n+\alpha) \right)
\nonumber \\
& & + \frac{N^2g^2}{32\pi^2} \int_0^1 {\rm d}\alpha \int_0^1 {\rm d}
\alpha' \sum_{n,n'} \frac{1-\cos(\theta(n+\alpha)-\theta(n-n'+\alpha'))}
{(n'-\alpha'+\alpha)^2} \nonumber \\
& = & -N\int \frac{{\rm d}p}{2\pi}\left( p \cos \theta_{\rm cont}(p)
+m \sin \theta_{\rm cont}(p) \right)
\nonumber \\
& & + \frac{N^2g^2}{8} \int \frac{{\rm d}p}{2\pi}\int \frac{{\rm d}p'}
{2\pi} \frac{1-\cos (\theta_{\rm cont}(p)-\theta_{\rm cont}(p'))}
{(p-p')^2} \ .
\label{t11}
\end{eqnarray}
Since the right hand side is $L$-independent, the pressure will be
equal to the vacuum pressure at all temperatures, to leading order in $1/N$.
Hence, there is no observable pressure of order
$N$, as one would expect from a confined system of quarks.
Treating the quark condensate along the same lines, we find
\begin{eqnarray}
\langle \bar{\psi}\psi \rangle & = & -\frac{1}{L} \sum_{n,i} \sin
\theta_i(n) \nonumber \\
& = & - \frac{N}{L} \sum_n \int_0^1 {\rm d} \alpha \sin \theta (n+\alpha)
\nonumber \\
& = & -N \int \frac{{\rm d}p}{2\pi} \sin \theta_{\rm cont}(p) \ .
\label{t12}
\end{eqnarray}
Once again, the sum over the discrete momenta and the color sum in
the large $N$ limit conspire to produce the continuum result, independently
of the starting $L$ value. In the alternative thermodynamic view,
the condensate does not depend on temperature to leading order
in $1/N$.
This leaves no room for a chiral phase transition,
not even in the limit $T\to \infty$. Finally, consider the expectation value
of the Polyakov loop. Due to our classical treatment and the assumed
field configuration, we trivially get zero,
\begin{equation}
\langle {\cal P} \rangle = \frac{1}{N}\sum_i {\rm e}^{{\rm i}\varphi_i}
\to \int_0^{2\pi} \frac{{\rm d}\varphi}{2\pi} {\rm e}^{{\rm i}\varphi}
= 0\ .
\label{t13}
\end{equation}
After interchanging space and time,
this signals confinement of static
charges at any temperature \cite{Svetitsky}. Physically, one would expect
screening by dynamical quarks, but this
cannot be seen yet in leading order in the $1/N$ expansion.
In the Wilson loop calculation for instance, screening by dynamical quarks
involves  at least one additional fermion loop. As is well
known, such diagrams are suppressed as compared to planar gluonic
diagrams by a factor $1/N$. A similar argument holds for the Polyakov loop.

Notice that so far, we have discussed the influence of the gauge
fields on the
quarks which is indeed dramatic. Vice versa, we do not expect the quarks
to influence significantly the gluon zero point motion, again
due to the effects of the Jacobian. The kinetic energy
$H_{\rm g}$ will give the same result as in pure Yang Mills theory.
Since this contribution to the energy density is $L$-independent, it
again yields
zero pressure,
reflecting the absence of physical gluonic excitations in 1+1 dimensions.

We now comment on the various studies of the finite temperature 't~Hooft
model mentioned in the beginning. The mechanism which we propose here has
some similarity with the color singlet projection in the partition
function as discussed by McLerran and Sen \cite{McLerran-Sen}, in
particular concerning the
$N$-dependence of the thermodynamic potential.
In Ming Li's calculation \cite{Ming Li}, the
problem seems to be the quark self energy. In Ref. \cite{LTLY}, it was
found that the principal value regulated self energy should be supplemented
by an additive constant $Ng^2L/48$ which diverges in the limit $L\to \infty$.
This
constant drops out of the calculation of color singlet mesons, but cannot
be simply ignored at the Hartree Fock level (the principal
value prescription makes sense for a first order pole, but not for a second
order pole). If we included this infinite constant into Ming
Li's calculation (or, equivalently, a finite temperature Hartree Fock
calculation), all the thermal factors would trivially vanish and we would
also get zero pressure and a $T$-independent condensate. In this
sense, the finite $L$ calculation presented above and the conventional
thermodynamic calculation are fully consistent.
Similar arguments
have already been put forward heuristically in ref. \cite{McLerran-Sen}.

As discussed above,
the problem with the chiral phase transition seen in ref. \cite{LTLY}
is the neglect of Jacobian and consequently freezing of gluons at the
point $\varphi_i=\pi$. The finite $L$ version of the theory solved there
is apparently not the gauge theory QCD$_2$, but has to be interpreted as
some other interacting fermion theory of Gross-Neveu type with a
Coulomb potential instead of a contact interaction. It cannot do full
justice to the confinement of quarks.
Finally, in \cite{DLMW}, it has tacitly been assumed that
the Jacobian can be accounted for by treating gluons as non-relativistic
fermions. This is certainly true for pure Yang Mills theory
where the idea originally came up \cite{Minahan-Poly}. In the
presence of quarks, the Hamiltonian is not invariant under permutations
of the $\varphi_i$. As can be seen from eqs. (\ref{p1}--\ref{p4}), it is
only invariant if one simultaneously permutes the quark colors --- this is
a residual gauge transformation.
Antisymmetrization of the gluon variables alone is thus not compatible
with the
time evolution of the system and cannot be used to satisfy the
boundary condition (\ref{p5a}).
The other point in which we differ from ref. \cite{DLMW} is the
issue of decompactification: In the
Schwinger model on the circle, the pure gauge field is a particle on a circle.
This circle
gets decompactified by the fermions --- the axial charge provides the
missing integer part of the coordinate \cite{Manton}. For reasons
discussed in \cite{Engelhardt},
such a phenomenon is hard to conceive in
the SU($N$) case, unlike what has been
put forward in \cite{DLMW}. Interestingly, we have found exactly the opposite
effect: The gluons decompactify the spatial circle on which the fermions
live, just because they are so much constrained by their own, compact
configuration space. This seems to be the mechanism by which a confined
system of quarks in 1+1 dimensions avoids wrong thermodynamic
behaviour to leading order in $1/N$.

We should like to thank M. Engelhardt for a helpful discussion and
M. Shifman for his interest in our work.

\end{document}